# Structural Disorder and Properties of the Stuffed Pyrochlore Ho$_2$TiO$_5$

G. C. Lau,[1] B.G. Ueland,[2] M. L. Dahlberg,[2] R.S. Freitas,[2,5] Q. Huang,[3] H. W. Zandbergen,[4] P. Schiffer[2] and R. J. Cava[1]

[1]Department of Chemistry, Princeton University, Princeton, New Jersey 08544, USA

[2]Department of Physics and Materials Research Institute, Pennsylvania State University, University Park, Pennsylvania 16802, USA

[3]NIST Center for Neutron Research, NIST, Gaithersburg, Maryland 20899, USA

[4]Department of Materials Science, Delft University of Technology, Rotterdamseweg 137, 2682 AL Delft, The Netherlands

[5]Present address: Instituto de Fisica, Universidade de Sao Paulo, C.P. 66318, Sao Paulo, SP 05315-970, Brazil.

**Abstract**

We report a structural and thermodynamic study of the "stuffed spin ice" material Ho$_2$TiO$_5$ (i.e., Ho$_2$(Ti$_{1.33}$Ho$_{0.67}$)O$_{6.67}$), comparing samples synthesized through two different routes. Neutron powder diffraction and electron diffraction reveal that the previously reported defect fluorite phase has short-range pyrochlore ordering, in that there are domains in which the Ho and Ho/Ti sublattices are distinct. By contrast, a sample prepared through a floating zone method has long range ordering of these sublattices. Despite the differences in crystal structures, the two versions of Ho$_2$TiO$_5$ display similar magnetic susceptibilities. Field dependent magnetization and measured recovered entropies, however, show a difference between the two forms, suggesting that the magnetic properties of the stuffed pyrochlores depend on the local structure.

**Introduction**

Pyrochlore compounds, with general formula $A_2B_2O_7$, represent an important family of materials that display geometrically frustrated magnetism. The frustrated geometry arises from sublattices of corner sharing tetrahedra, which are present for both the *A* and *B* cations. Spin ice pyrochlores ($Ln_2M_2O_7$, where *Ln* = Dy, Ho and *M* = Ti, Sn) have been of considerable interest and studied extensively[1-9] as unique examples of magnetic frustration where the spins have effective ferromagnetic interactions. The geometry of the rare earth sublattice, combined with crystal field effects that restrict the moments to be Ising-like, generate spin frustration that mimics the positional frustration of hydrogen atoms in water ice [8-10]: spins or hydrogen atoms sitting on the corners of tetrahedra seek a minimum energy configuration in the short range by freezing into a 'two-in, two-out' arrangement.[1] The large degeneracy that arises based on energetically equivalent arrangements of two-in and two-out on a single tetrahedron leads to overall long range disorder. The same measurable zero-point entropy for both spin ice[9] and water ice[11, 12] exists according to the 'ice rules,'[13, 14] and is directly attributable to this degeneracy.

Dilution studies on spin ice materials,[15-17] where magnetic rare earth moments are replaced with a nonmagnetic species, reveal that decreasing the spin interactions does not destroy the ice-like state, but does suppress the magnitude of the freezing signature. It was recently shown that 'stuffed spin ice' - the opposite case where additional magnetic atoms are stuffed into the nonmagnetic Ti sites creating, for example, the series $Ho_2(Ti_{2-x}Ho_x)O_{7-x/2}$, $0 \leq x \leq 0.67$ - retains the same zero-point entropy as undoped spin ice and may possess accelerated spin dynamics.[18]

$Ho_2(Ti_{2-x}Ho_x)O_{7-x/2}$ represents a continuous solid solution from $Ho_2Ti_2O_7$ (x=0) to $Ho_2(Ti_{1.33}Ho_{0.67})O_{6.67}$ (x=0.67), or equivalently, $Ho_2TiO_5$.[19-24] The extra Ho in stuffed spin ice replaces Ti for small x, i.e., the excess Ho are located on the *B*-site sublattice of the pyrochlore structure. The *B*-sites form a sublattice of corner-sharing tetrahedra equal in size and atomic distances to the *A*-site sublattice but offset by (½, ½, ½). At lower doping levels, $0 \leq x \leq 0.3$, the structure retains this pyrochlore ordering with a clear distinction between the *A* and *B* sublattices. The extra Ho is confined primarily to the *B*-site, and the *A* sublattice remains largely undisturbed. As more Ho is substituted in place of Ti (x > 0.3), some Ti begins mixing onto the Ho *A*-site, and both *A* and *B*-sites have mixed occupancy at x=0.67 doping.[19] The average structure is of the fluorite type, where the *A* and *B* cations are randomly mixed on metal sites. This transformation turns the magnetic lattice from corner sharing tetrahedra in the pyrochlore $Ho_2Ti_2O_7$, to edge sharing tetrahedra in the fluorite $Ho_2TiO_5$.

Here we report a comparison of the structure and magnetic properties of $Ho_2TiO_5$ synthesized in two different ways. The first synthesis condition involves firing the starting materials to a high temperature followed by a rapid quench to room temperature. The average structure and magnetic properties of this variant have been reported previously.[18, 19] This material is referred to as quenched (Q) $Ho_2TiO_5$. The second synthesis condition involves making $Ho_2TiO_5$ with a floating zone crystal growth technique, where the material is melted and cooled at a much slower rate than that used in the Q synthesis. This material is referred to as floating zone (FZ) $Ho_2TiO_5$.

The Q sample displays on average a complete disordering of the Ho and Ti cations to form the fluorite structure.[18, 19] Here we report neutron powder diffraction and electron diffraction data that give evidence that short range pyrochlore ordering exists in this highly disordered material. This ordering takes the form of small pyrochlore domains in which the extra stuffed Ho mixes in a disordered fashion into the *B*-site sublattice, while the *A*-site sublattice, containing the usual pyrochlore Ho arrangement, is largely undisturbed. The domains have short correlation lengths, which we postulate to arise from an antiphase arrangement of the pyrochlore regions. These antiphase domains are postulated to form on cooling from a fully disordered cation array at high temperature, nucleating into pyrochlore-like domains where one or the other of the interpenetrating tetrahedral sublattices is selected locally to be the "A" site in the short range ordered pyrochlore. The average over these small pyrochlore domains appears as a disordered fluorite structure, as reported earlier.

In contrast to the Q sample, we find the FZ sample to be a long range ordered pyrochlore phase, with the A lattice consisting almost entirely of Ho ions and the extra Ho mixing primarily on the Ti *B*-site. Here we show that the distinction in the long range versus short range cation ordering does not result in significant differences in magnetic properties between the FZ and Q variants. The recovered entropy measured for the FZ sample is, however, greater than that expected for ice-like materials, while the missing entropy of the Q version remains similar to that present in ordinary $Ho_2Ti_2O_7$ spin ice.

**Experimental**

$Ho_2(Ti_{1.33}Ho_{0.67})O_{6.67}$, or $Ho_2TiO_5$, was prepared in two different ways. In both cases, $Ho_2O_3$ (Cerac, 99.9%) and $TiO_2$ (Cerac, 99.9%) powders were thoroughly mixed in a 1:1 molar ratio with an agate mortar and pestle. For the quenched sample, the powders were pressed into a pellet, wrapped in molybdenum foil, heated at 1700 °C in a static argon atmosphere for 12 hours, and quenched to room temperature in approximately 30 minutes. The argon atmosphere was achieved in a vacuum furnace first evacuated to about $10^{-6}$ torr and then back-filled with argon (Airgas, 99.9%) to room pressure. The floating zone sample required the same initial treatment as the quenched version to ensure chemical homogeneity. The sintered pellet was reground to a fine powder and formed into cylindrical rods in a sealed rubber tube pressed for 15 minutes at 70 MPa in a cold isostatic press. The polycrystalline rods were sintered in air at 1400 °C for 12 h before use in a Crystal Systems optical image floating zone furnace. The crystal was grown in flowing air at rates between 2.00 and 10.00 mm/h, and pulverized afterwards for structure and physical properties characterization.

Both samples were analyzed for phase purity by powder X-ray diffraction (XRD) using Cu *Kα* radiation and a diffracted beam graphite monochromator. Neutron diffraction (ND) data were collected on both samples at the NIST Center for Neutron Research on the high resolution powder neutron diffractometer with monochromatic neutrons of wavelength 1.5403 Å produced by a Cu(311) monochromator. Collimators with horizontal divergences of 15′ and 20′of arc were used before and after the monochromator, and a collimator with a horizontal divergence of 7′ was used after the sample. Data were collected in the 2*θ* range of 3°–168° with a step size of 0.05°. Rietveld refinements of the structures were performed with the GSAS suite of programs.[25] The peak shape was described with a pseudo-Voigt function. The background was fit to 12 terms in a linear interpolation function. The neutron scattering amplitudes used in the refinements were 0.801, -0.344, and 0.580 ($\times 10^{-12}$ cm) for Ho, Ti, and O, respectively.

Electron microscopy analysis was performed with a Philips CM200 electron microscope having a field emission gun and operated at 200 kV. Electron-transparent areas of specimens were obtained by crushing them slightly under ethanol to form a suspension and then by dripping a droplet of this suspension on a carbon-coated holey film on a Cu or Au grid.

Magnetic and specific heat measurements were performed on pressed pellets in Quantum Design MPMS and PPMS cryostats. The magnetizations of the samples were measured down to T = 1.8 K and in fields up to H = 7 T. Fits to the Curie-Weiss law were performed to the DC susceptibility $\chi$ = M/H, using magnetization data taken at H = 0.1 T. Heat capacity measurements were performed using a standard semi-adiabatic heat pulse technique, and the addendum heat capacity was measured separately and subtracted. The samples used for heat capacity measurements were thoroughly mixed with Ag powder before being pressed into a pellet, to facilitate thermal conductivity throughout the sample. The contribution to the heat capacity from the Ag was subtracted off using previously published data.[26] All the samples for susceptibility were cut to needle-like shapes, and the long side was oriented along the direction of the applied field, in order to minimize demagnetization effects.

**Results and Discussion**

Rietveld refinement fits to the neutron powder diffraction (ND) data are shown in Figure 1 to contrast the difference in crystal structures of $Ho_2TiO_5$ made using the two methods. The patterns reveal that both long range and short range order are present in the materials, seen in the presence of both narrow and broad diffraction peaks. The fluorite structure for $Ho_2TiO_5$ is face centered cubic, with $a \approx 5.15$Å. The pyrochlore structure, in contrast, is face centered cubic with $a \approx 10.3$Å, a 2x2x2 supercell of the fluorite structure due to the distinction between the *A* and *B* sites. Thus the diffraction patterns consist of a series of peaks from the fluorite-like structure (fluorite substructure peaks) with additional peaks (the pyrochlore superstructure peaks) that appear with increasing intensity as the pyrochlore-type ordering becomes more developed. If the pyrochlore ordering occurs over only a short range, then the pyrochlore superstructure peaks will be broadened. This is seen in Figure 1. The neutron data used in the refinements were taken at room temperature. A diffraction pattern of the Q sample at 4 K was virtually identical to the room temperature data, indicating the structure has no temperature dependence. In the present study, the significantly broadened superstructure peaks were omitted from the refinements, as a detailed analysis of neutron powder patterns in the $Ho_2(Ti_{2-x}Ho_x)O_{7-x/2}$ series, including fits to the broadened peaks, will be published elsewhere.[27] For the Q sample, all of the pyrochlore superstructure peaks were broadened. For the FZ sample, only the regions near the 331 and 422 pyrochlore peaks were excluded from the refinement. The data for both Q and FZ samples were refined within a pyrochlore structure model to better compare lattice parameters and cation site occupancies. Ho and Ti occupancies were allowed to refine freely on both the *A* and *B*-sites of the pyrochlore structure, with the constraint that the occupancies summed to 1 on each site (i.e. all metal sites are fully occupied). The oxygen occupancies were also allowed to refine freely. Thermal displacement parameters were constrained for cations mixed on the same site and for oxygen atoms in the similar 8b and 8a sites. In general, all of the atoms gave relatively large thermal displacement parameters in both samples when allowed to refine freely. This is due to the high degree of average positional disorder seen in the cation and oxygen lattices of both phases. The results from the refinements are displayed in Table I.

The Q sample displays average long range disorder between Ho and Ti atoms. This is evidenced by the presence of sharp, well defined fluorite subcell peaks and the lack of equally sharp pyrochlore superstructure peaks. Considering only the sharp fluorite peaks, the refinement shows that Ho and Ti are randomly mixed in an approximately 2:1 ratio on both 16d and 16c cation sites, as expected from the material stoichiometry. However, the pyrochlore superstructure peaks are not entirely absent and are actually broadened. This is most apparent for the 331 peak, the highest intensity pyrochlore supercell reflection, and is true for all of the pyrochlore superstructure reflections. The broadened peaks show that there remains short range pyrochlore-like ordering in the Q sample. These broad peaks were not seen in the X-ray diffraction data (not shown) or mentioned in previous structural reports on $Ho_2TiO_5$, which employed X-ray diffraction.[18, 19, 22-24] This indicates that the oxygen atoms, which are relatively transparent to X-rays, may contribute to the short range order scattering. In $Ho_2Ti_2O_7$ and other ordered pyrochlores, the oxygen atoms fully occupy 8b and 48f sites while the 8a site is entirely vacant.[28] In the fluorite structure, the *A* and *B* atoms are mixed and no longer distinguishable, and the oxygen atoms may occupy any of the 8b, 8a, or 48f sites.[28] The structure refinement reported here on the Q sample shows that the Ho-stuffing alters the oxygen lattice by introducing occupancy onto the 8a site while decreasing the amount of oxygen on the 48f site. In an ideal fluorite structure, the x position of the 48f oxygen is 0.375. In pyrochlores, the x position shifts towards 0.3125 to stabilize the ordering of the cations. The refinement on the Q sample gives the x position of the 48f oxygen to be 0.368, close to the ideal fluorite value.

The FZ sample contrasts with the Q phase by displaying sharp, long range ordered peaks that are well described by a cubic pyrochlore model, although some diffuse scattering is still observed, particularly around the 331 and 422 peaks. Despite omitting those peaks from the refinement, the thermal displacement parameter of the 16c site refined to an unusually large value. However, by fixing the $U_{iso}$ to a reasonable value, the quality of fit did not change significantly, and the site occupations of the cations changed by less than 5%. We attribute this to the low intensity broad reflections present throughout the data. Small regions of short range ordered pyrochlore, as in the Q phase, could contribute to this scattering. However, a variety of low intensity peaks were not well fit by a simple pyrochlore model (see insets of Figure 1), suggesting the presence of an additional structural modulation, as described further below. As can be seen in Table I, refinement shows that the extra stuffed Ho in the FZ version mixes primarily on the Ti *B*-site, leaving the original Ho site largely unaffected. Figure 2 shows the average crystal structures of the cation sublattices in both Q and FZ $Ho_2TiO_5$, and gives approximate cation occupancies. The oxygen lattice of the FZ phase shows partial occupancies in all three oxygen sites. The x position of the 48f oxygen is 0.349, shifted away from the ideal fluorite value.

Electron diffraction patterns (EDP) comparing the Q and FZ samples are given in Figure 3, which shows projections along the <110> zone axis. The strongest spots in both cases represent the underlying fluorite sub-cell common to both structures. The Q EDP in Figure 3a displays superreflections situated halfway in between the fluorite spots (e.g. 111, 331 referred to the pyrochlore cell) providing additional evidence of pyrochlore ordering in the sample. These pyrochlore reflections are greatly elongated in the 111 direction, however, indicating that the ordering is on a short length scale, consistent with the broadening of the pyrochlore superstructure peaks in the neutron diffraction data. This structure model, which describes short range pyrochlore ordering within an average disordered fluorite, is similar to that observed by electron diffraction for cubic stabilized zirconia materials.[29-37] Given the large size difference between Ho and Ti, it is not surprising to find that despite the complete disorder of the average structure seen in the Q defect fluorite over long length scales, ordering still occurs on the local scale. Indeed, when allowed to cool from high temperature at a slower rate, as in the FZ sample, the ordering occurs over a longer range, yielding the sharper pyrochlore reflections seen in Figure 3b. Only a slight elongation of these spots exists along the 111 direction (the 111 peak in the neutron powder diffraction data is also slightly broadened). This indicates that the domains of pyrochlore ordering in the FZ material are on average much larger than the pyrochlore domains in the Q material.

The additional weak reflections seen in the EDP show that the details of the pyrochlore ordering in both samples are actually more complicated. Additional weak reflections of this type indicate the presence of a minor structural modulation of the cubic pyrochlore structure in both cases. These reflections can be described as a 7-fold increase of the pyrochlore unit cell in the 662 direction. This 7x supercell is observed in both types of materials (Figure 3). An additional tripling of the pyrochlore unit cell in the 111 direction is seen only in the FZ sample. These reflections are also elongated, indicating the tripling occurs only in the short range. Nanodiffraction (spot size about 5 nm) and HREM imaging show that the diffraction patterns in Figure 3 are a composite of diffraction patterns from regions in which only one superstructure is present. The complexity of the modulations implied by the distribution of the supercell reflections makes detailed structural characterization of this modulation beyond the scope of the present study. The weak diffraction from the 7x and 3x superstructures in the modulated phase in the FZ ND pattern is likely responsible for the small peaks not indexed in the inset of Figure 1b, explaining the relatively larger weighted residuals of the FZ Rietveld fit.

We surmise that antiphase domains are responsible for the short range pyrochlore ordering described above. The Ho and Ti atoms in $Ho_2(Ti_{1.33}Ho_{0.67})O_{6.67}$ are randomly mixed on both the *A* and *B* pyrochlore sites at high temperature. As the material is quenched, Ho and Ti will naturally try to order separately from one another due to their large difference in size. This results in pure Ho nucleating out on one of the two interpenetrating sublattices of corner sharing tetrahedra: this becomes the *A*-site of the pyrochlore ordering in this local region of the material. The extra stuffed Ho is then forced onto the other pyrochlore cation site and mixed with the remaining Ti. The antisite domains, where one or the other of the interpenetrating sublattices is chosen locally by Ho, are frozen in by the quench, and short range pyrochlore ordering is the consequence. Taken as a powder average, the Ho and Ti appear disordered over both *A* and *B* sites resulting in an average fluorite structure. The FZ process allows the material to cool slowly and the antisite domains can anneal into just one type of site ordering. This is supported by the long range sharp pyrochlore peaks in the neutron pattern and the sharp pyrochlore spots in the EDP.

The magnetic susceptibility of the FZ sample is given in Figure 4, showing no difference between the zero-field cooled and field cooled data. The lack of a bifurcation precludes the material from being a spin glass in the temperature regime studied, similar to the behavior reported previously for the Q sample.[18] A $\chi^{-1}(T)$ comparison between Q and FZ is displayed in the inset, with the Curie Weiss temperatures ($\theta_w$) and effective magnetic moments (*p*) determined from fits to both the high temperature (50-150 K) and low temperature data (10-20 K) displayed in Table II. Both samples show no long-range magnetic order down to 2 K, and have similar negative $\theta_w$'s, which indicate dominant antiferromagnetic spin interactions. This is in contrast to undoped $Ho_2Ti_2O_7$ spin ice, which has weakly ferromagnetic interactions.[38] The determined *p* values are similar in both Q and FZ samples also, and are close to the expected value for a free $Ho^{3+}$ ion. The differences in long range vs. short range ordering between the Q and FZ samples therefore do not result in significant differences in the magnetic susceptibility.

Figure 5 compares the field dependence of the magnetization for both the Q and FZ samples. Although they both reach similar magnetizations at 6 Tesla, the Q sample begins to saturate at a lower applied field. The FZ phase requires a stronger applied field before approaching the same magnetization, and appears would saturate at a higher magnetization if extrapolated to stronger fields. This is consistent with the slightly more negative $\theta_w$ in the FZ sample, indicating slightly stronger antiferromagnetic interactions that would be harder to saturate with an external field. The M(H=6 T) values are given in Table II, and are approximately half of the full magnetization expected for a free $Ho^{3+}$ ion. In $Ho_2Ti_2O_7$, the magnetization saturates at half as well[38] suggesting that despite the difference in average spin connectivity between $Ho_2Ti_2O_7$ and the Q and FZ $Ho_2TiO_5$, their local single-ion anisotropies may be similar.

Specific heat data, C(T), for the FZ and Q samples are plotted in Figure 6a. Both samples show a magnetic peak around T = 1.6 K, followed by the beginning of a lower T Schottky peak, due to the hyperfine contributions from the Ho nuclei. The peak at T=1.6 K is much sharper in the FZ sample, suggesting that a longer range ordering is occurring in this material. Figure 6b compares the magnetic entropy, S(T), at H = 0 of the FZ and Q samples, determined by first subtracting the lattice and nuclear spin contributions from the total specific heat followed by integrating $C_{magnetic}(T)/T$ from low to high T. As reported previously,[18] the magnetic entropy of the Q sample remains ice-like, saturating at R[ln2 – 1/2ln(3/2)] rather than the expected entropy for a two-state system, Rln2. This was unexpected as the average structure of the Q sample consists of an array of edge-sharing tetrahedra of Ho atoms, contrary to the corner-sharing tetrahedral arrangement in $Ho_2Ti_2O_7$. The recovered magnetic entropy of the FZ sample is shown to be larger than that of the Q sample. Because of the differences in average crystal structure, a disparity in the measured entropy is not surprising.

**Conclusions**

The synthesis of a long range ordered pyrochlore-like phase of $Ho_2TiO_5$ using the floating zone method is reported. The structure and properties of this FZ sample are compared with the previously reported fluorite-like Q phase of $Ho_2TiO_5$. The Q sample is actually pyrochlore-like on the local scale, where the extra Ho mixes primarily onto the Ti $B$-site while the $A$-site remains undisturbed. The FZ sample displays this pyrochlore-like ordering in the long range. Both materials exhibit an additional 7-fold structural modulation, but only the FZ phase shows another 3-fold modulation in the 111 direction. While the differences in structure of the two variants do not affect the magnetic susceptibility significantly, the field dependent magnetization and magnetic entropy are clearly influenced. One possibility is that the extra 3-fold structural modulation observed in the FZ phase of $Ho_2TiO_5$ breaks the cubic symmetry of the pyrochlore lattice significantly, resulting in our observation of a lower ground state entropy for that variant. The structural information reported here is important for the modeling of the magnetic behavior of stuffed spin ice: since we demonstrate that the "stuffed" Ho in these spin ices goes on the pyrochlore B sites, locally adding magnetic neighbors to an undisturbed pyrochlore lattice. Similar studies of other stuffed rare earth pyrochlores will lend insight into whether these structural observations can be generalized to the wide range of frustrated magnetic materials in this family.

**Acknowledgements**

This work was supported by the National Science Foundation Division of Materials Research (DMR-0353610). R.S.F. thanks the CNPq-Brazil for support. Certain commercial chemicals and equipment are identified in this report to describe the subject adequately. Such identification does not imply recommendation or endorsement by the NIST, nor does it imply that the equipment identified is necessarily the best available for the purpose.

**References**

[1] S. T. Bramwell and M. J. P. Gingras, Science **294**, 1495 (2001).

[2] J. E. Greedan, Journal Of Alloys And Compounds **408**, 444 (2006).

[3] J. Snyder, B. G. Ueland, J. S. Slusky, H. Karunadasa, R. J. Cava, and P. Schiffer, Physical Review B **69**, 064414 (2004).

[4] R. Higashinaka, H. Fukazawa, and Y. Maeno, Physical Review B **68**, 014415 (2003).

[5] H. Fukazawa, R. G. Melko, R. Higashinaka, Y. Maeno, and M. J. P. Gingras, Physical Review B **65**, 054410 (2002).

[6] S. T. Bramwell, M. J. Harris, B. C. den Hertog, M. J. P. Gingras, J. S. Gardner, D. F. McMorrow, A. R. Wildes, A. L. Cornelius, J. D. M. Champion, R. G. Melko, and T. Fennell, Physical Review Letters **87**, 047205 (2001).

[7] K. Matsuhira, Y. Hinatsu, K. Tenya, and T. Sakakibara, Journal Of Physics-Condensed Matter **12**, L649 (2000).

[8] M. J. Harris, S. T. Bramwell, D. F. McMorrow, T. Zeiske, and K. W. Godfrey, Physical Review Letters **79**, 2554 (1997).

[9] A. P. Ramirez, A. Hayashi, R. J. Cava, R. Siddharthan, and B. S. Shastry, Nature **399**, 333 (1999).

[10] S. T. Bramwell and M. J. Harris, Journal Of Physics-Condensed Matter **10**, L215 (1998).

[11] W. F. Giauque and J. W. Stout, Journal Of The American Chemical Society **58**, 1144 (1936).

[12] W. F. Giauque and M. F. Ashley, Physical Review **43**, 81 (1933).

[13] J. D. Bernal and R. H. Fowler, Journal Of Chemical Physics **1**, 515 (1933).

[14] L. Pauling, Journal Of The American Chemical Society **57**, 2680 (1935).

[15] G. Ehlers, J. S. Gardner, C. H. Booth, M. Daniel, K. C. Kam, A. K. Cheetham, D. Antonio, H. E. Brooks, A. L. Cornelius, S. T. Bramwell, J. Lago, W. Haussler, and N. Rosov, Physical Review B **73**, 174429 (2006).

[16] J. Snyder, J. S. Slusky, R. J. Cava, and P. Schiffer, Physical Review B **66**, 064432 (2002).

[17] J. Snyder, B. G. Ueland, A. Mizel, J. S. Slusky, H. Karunadasa, R. J. Cava, and P. Schiffer, Physical Review B **70**, 184431 (2004).

[18] G. C. Lau, R. S. Freitas, B. G. Ueland, B. D. Muegge, E. L. Duncan, P. Schiffer, and R. J. Cava, Nature Physics **2**, 249 (2006).

[19] G. C. Lau, B. D. Muegge, T. M. McQueen, E. L. Duncan, and R. J. Cava, Journal Of Solid State Chemistry **179**, 3126 (2006).

[20] C. E. Bamberger and G. M. Begun, Journal Of The Less-Common Metals **109**, 147 (1985).

[21] C. E. Bamberger, G. M. Begun, J. Brynestad, and H. W. Dunn, Inorganica Chimica Acta-F-Block Elements Articles And Letters **109**, 141 (1985).

[22] G. E. Sukhanova, K. N. Guseva, A. V. Kolesnikov, and L. G. Shcherbakova, Inorganic Materials **18**, 1742 (1982).

[23] M. A. Petrova, A. S. Novikova, and R. G. Grebenshchikov, Inorganic Materials **18**, 236 (1982).

[24] G. E. Sukhanova, K. N. Guseva, L. G. Mamsurova, and L. G. Shcherbakova, Inorganic Materials **17**, 759 (1981).

[25] A. Larson and R. C. Von Dreele, (Los Alamos National Laboratory, Los Alamos, NM, 1994).

[26] P. U. T. P. R. Center., in *Thermophysical properties of matter*, edited by Y. S. Touloukian (IFI/Plenum, New York, 1970), Vol.

[27] G. C. Lau, Q. Huang, T. M. McQueen, and R. J. Cava, In preparation (2007).

[28] M. A. Subramanian, G. Aravamudan, and G. V. S. Rao, Progress In Solid State Chemistry **15**, 55 (1983).

[29] D. N. Argyriou, M. M. Elcombe, and A. C. Larson, Journal Of Physics And Chemistry Of Solids **57**, 183 (1996).

[30] S. Garcia-Martin, M. A. Alario-Franco, D. P. Fagg, A. J. Feighery, and J. T. S. Irvine, Chemistry Of Materials **12**, 1729 (2000).

[31] S. Garcia-Martin, M. A. Alario-Franco, D. P. Fagg, and J. T. S. Irvine, Journal Of Materials Chemistry **15**, 1903 (2005).

[32] I. R. Gibson and J. T. S. Irvine, Journal Of Materials Chemistry **6**, 895 (1996).

[33] Y. Tabira, R. L. Withers, J. C. Barry, and L. Elcoro, Journal Of Solid State Chemistry **159**, 121 (2001).

[34] R. L. Withers, J. G. Thompson, and P. J. Barlow, Journal Of Solid State Chemistry **94**, 89 (1991).

[35] R. L. Withers, J. G. Thompson, P. J. Barlow, and J. C. Barry, Australian Journal Of Chemistry **45**, 1375 (1992).

[36] R. L. Withers, J. G. Thompson, N. Gabbitas, L. R. Wallenberg, and T. R. Welberry, Journal Of Solid State Chemistry **120**, 290 (1995).

[37] Y. Tabira, R. Withers, J. Thompson, and S. Schmid, Journal Of Solid State Chemistry **142**, 393 (1999).

[38] S. T. Bramwell, M. N. Field, M. J. Harris, and I. P. Parkin, Journal Of Physics-Condensed Matter **12**, 483 (2000).

TABLE I. Crystallographic data for $Ho_2TiO_5$ in the space group Fd-3m (no. 227). ($U_{iso}$ = isothermal temperature factor; Occ = occupancy.)

| Compound | Atom | Wyckoff position | x | y | z | $U_{iso}$*100 | Occ |
|---|---|---|---|---|---|---|---|
| $Ho_2TiO_5$ | Ho(1) | 16d | 0.5 | 0.5 | 0.5 | 1.6(4) | 0.65(3) |
| Quenched | Ti(1) | 16d | 0.5 | 0.5 | 0.5 | 1.6(4) | 0.35(3) |
|  | Ti(2) | 16c | 0 | 0 | 0 | 6.1(5) | 0.28(4) |
|  | Ho(2) | 16c | 0 | 0 | 0 | 6.1(5) | 0.72(4) |
|  | O(1) | 8b | 0.375 | 0.375 | 0.375 | 7.0(2) | 1.0(1) |
|  | O(2) | 8a | 0.125 | 0.125 | 0.125 | 7.0(2) | 0.73(1) |
|  | O(3) | 48f | 0.368(3) | 0.125 | 0.125 | 8.6(7) | 0.82(3) |

a (Å) = 10.3011(9)
$\chi^2$ = 1.05; Rwp = 5.16%; Rp = 4.02%.

| Compound | Atom | Wyckoff position | x | y | z | $U_{iso}$*100 | Occ |
|---|---|---|---|---|---|---|---|
| $Ho_2TiO_5$ | Ho(1) | 16d | 0.5 | 0.5 | 0.5 | 1.86(8) | 0.936(7) |
| Floating Zone | Ti(1) | 16d | 0.5 | 0.5 | 0.5 | 1.86(8) | 0.064(7) |
|  | Ti(2) | 16c | 0 | 0 | 0 | 1.8 | 0.685(4) |
|  | Ho(2) | 16c | 0 | 0 | 0 | 1.8 | 0.315(4) |
|  | O(1) | 8b | 0.375 | 0.375 | 0.375 | 2.0(2) | 0.79(3) |
|  | O(2) | 8a | 0.125 | 0.125 | 0.125 | 2.0(2) | 0.20(2) |
|  | O(3) | 48f | 0.3491(6) | 0.125 | 0.125 | 7.9(3) | 0.95(2) |

a (Å) = 10.2934(7)
$\chi^2$ = 2.67; Rwp = 6.39%; Rp = 4.63%.

TABLE II. Weiss constants and magnetic moments determined from the Curie-Weiss fits of high and low temperature portions of the magnetic susceptibilities.

|  | high temp (50-150 K) | | low temp (10-20 K) | | |
|---|---|---|---|---|---|
|  | $\theta_w$ (K) | p ($\mu_B$) | $\theta_w$ (K) | p ($\mu_B$) | M (H=6T) |
| $Ho_2TiO_5$ Q | -5.88 | 10.23 | -2.13 | 10.00 | 5.60 |
| $Ho_2TiO_5$ FZ | -5.92 | 10.40 | -3.36 | 10.24 | 5.46 |

**Figure Captions**

Figure 1. (Color online) Neutron powder diffraction data at room temperature (crosses), Rietveld refinement fits (green), and difference curves (black) are shown in the main panels for both the quenched (a) and floating zone (b) $Ho_2TiO_5$ samples. Regions in the plots with no calculated or difference curves represent the portions of the neutron data omitted from the refinements. The insets enlarge the region around the 331 peaks, revealing small diffuse peaks especially in the FZ phase.

Figure 2. (Color online) The average cation lattices of the Quenched (a) and Floating Zone (b) phases showing the approximate site occupancies.

Figure 3. <110> zone axis electron diffraction patterns are shown for the Q (a) and FZ (b) versions of $Ho_2TiO_5$. The reflections are indexed assuming a pyrochlore structure. Pyrochlore reflections are streaked in the Q phase, as evidenced by the 111 reflection. This indicates short range ordered pyrochlore regions. The FZ phase displays only a slight broadening of the 111 reflections indicating longer ranged pyrochlore ordering than in the Q material. Both Q and FZ phases display an additional complex modulation of the pyrochlore structure consisting of a 7-fold enlargement of the unit cell in the 662 direction. The FZ material also shows a 3-fold enlargement of the unit cell in the 111 direction.

Figure 4. (Color online) Main panel: Magnetic susceptibility as a function of temperature is plotted for the FZ sample only, showing both zero field cooled (ZFC) and field cooled (FC) data. Inset: The inverse susceptibility versus temperature is compared between Q and FZ variants.

Figure 5. (Color online) Field dependence of the magnetization at 2 K is displayed for both Q and FZ samples.

Figure 6. (Color online) Total specific heat (a) and total magnetic entropy (b) at zero applied field is presented for Q and FZ samples. The entropy is obtained by integrating the magnetic specific heat from below T = 1 to 22 K.

Figure 1.

a.

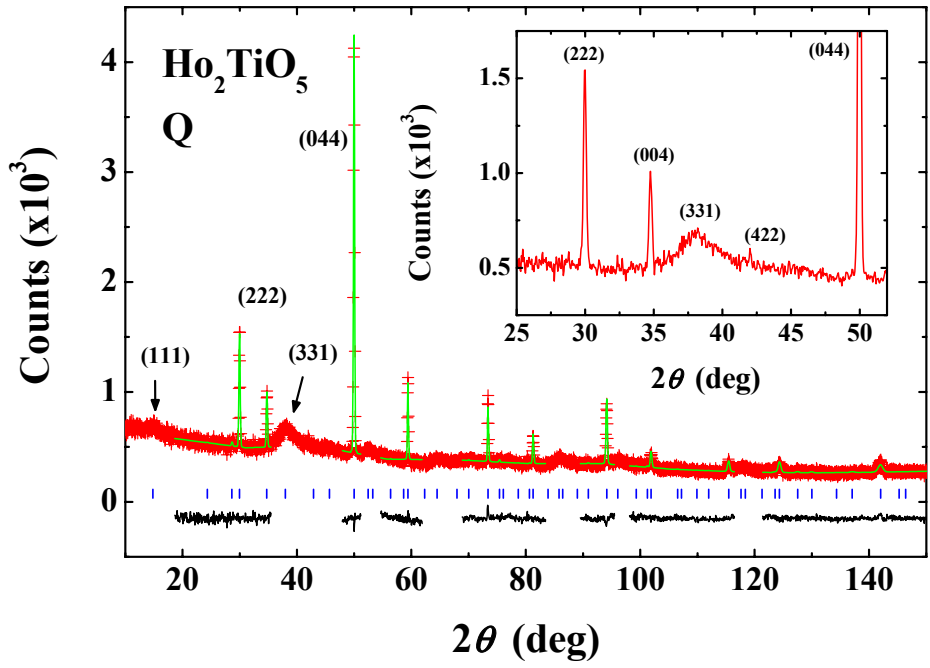

b.

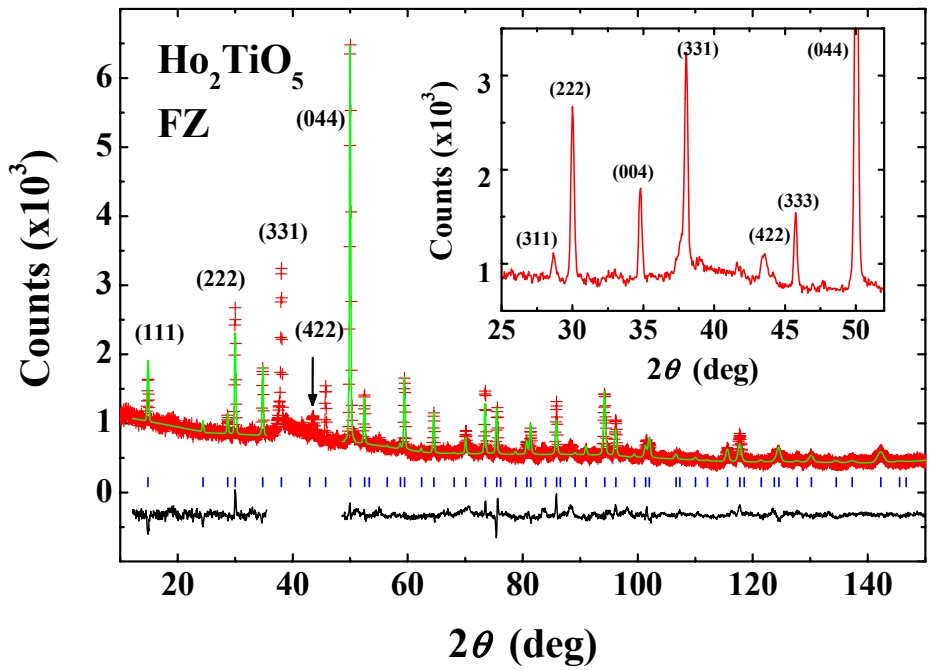

Figure 2.

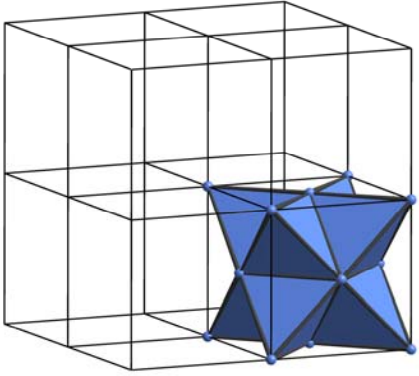 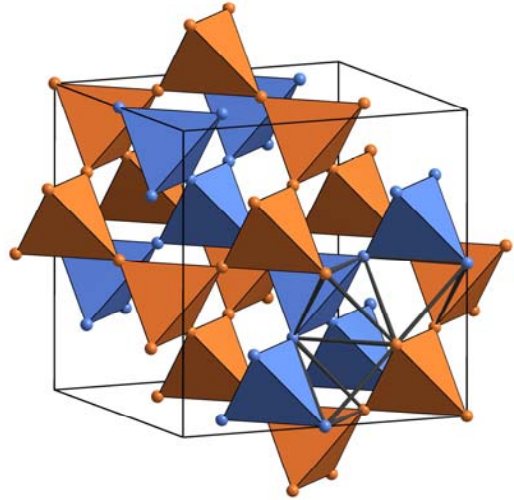

**Quenched**  **Floating Zone**

Figure 3.

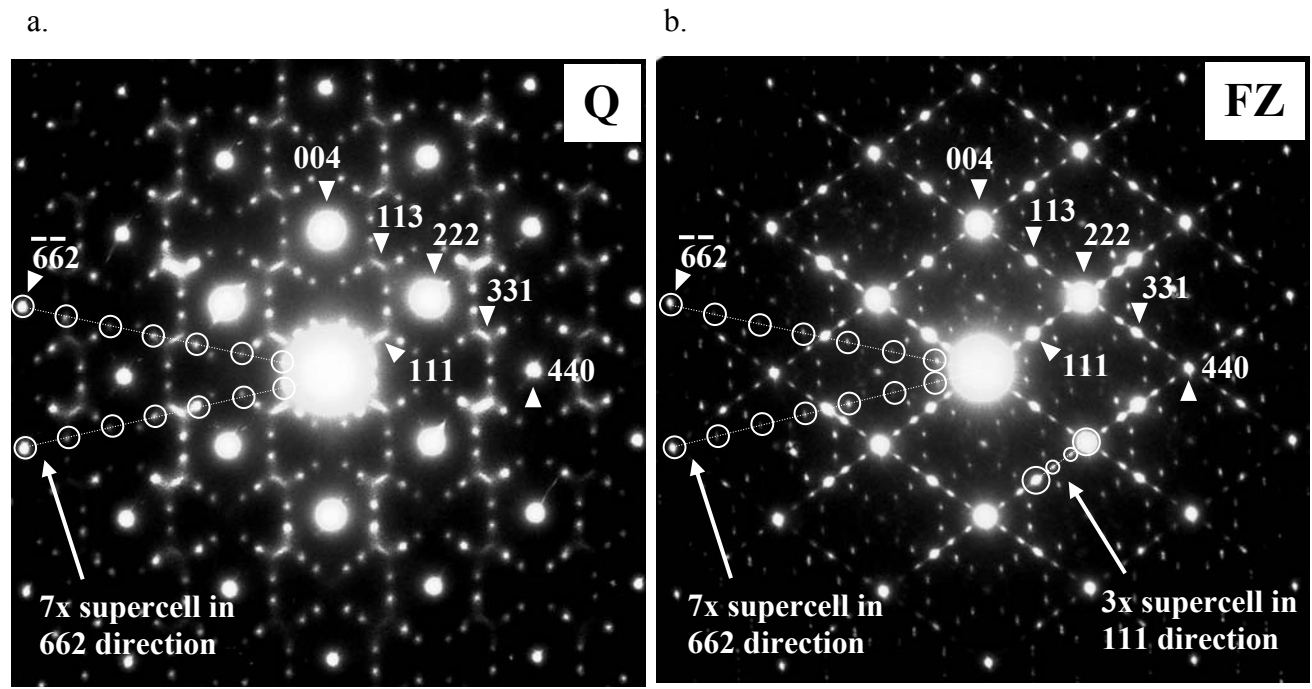

Figure 4.

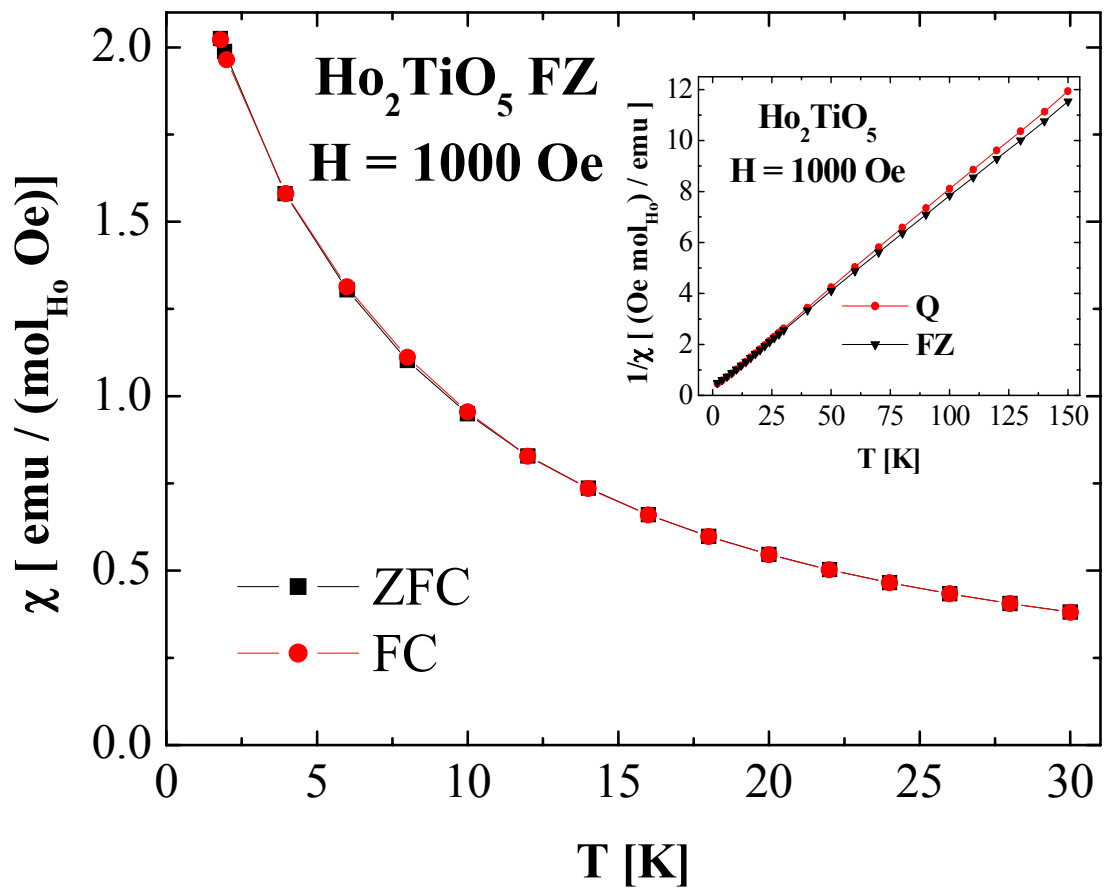

Figure 5.

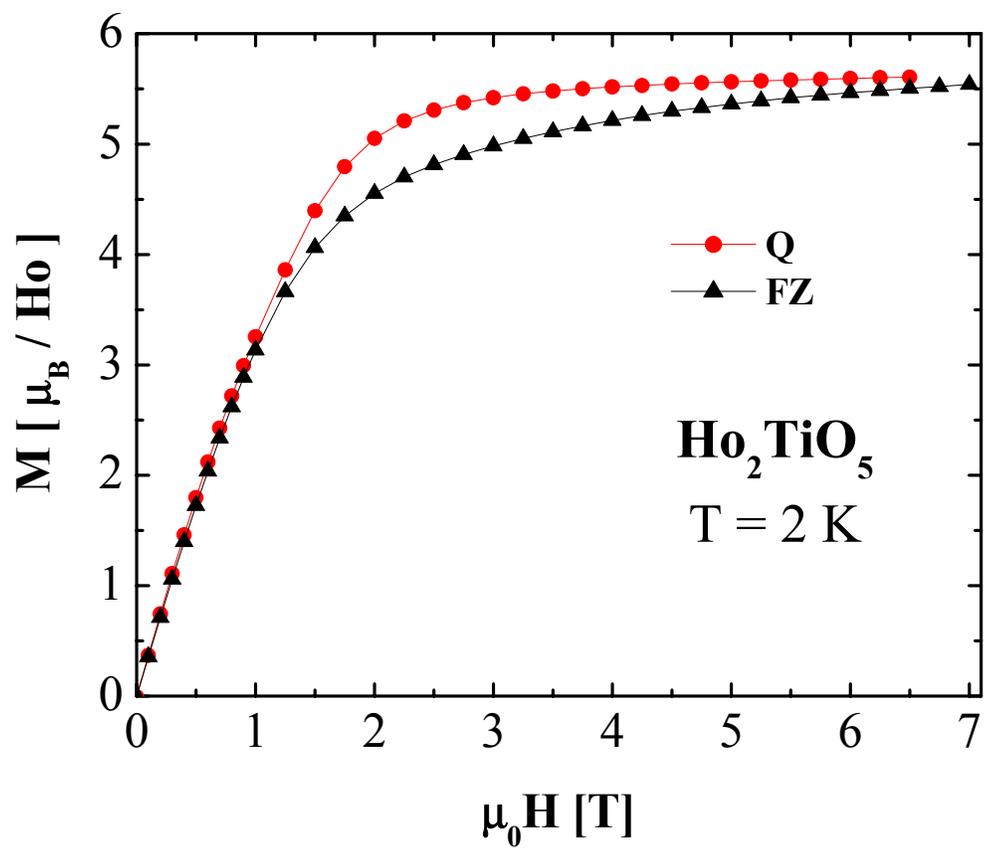

Figure 6.

a.

b.

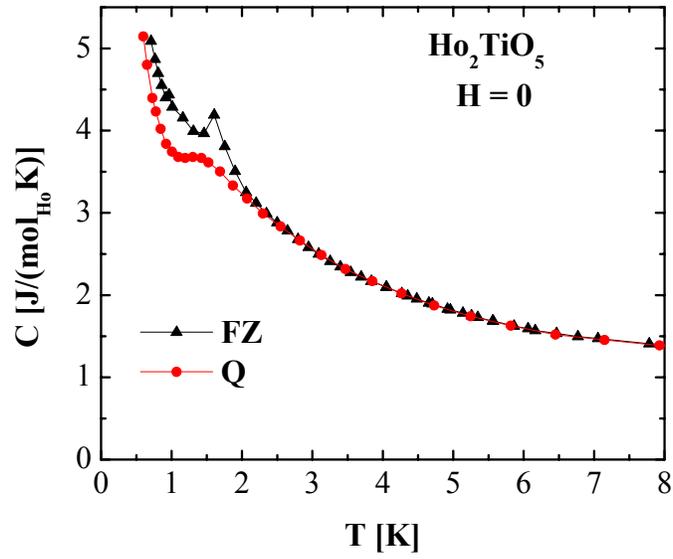
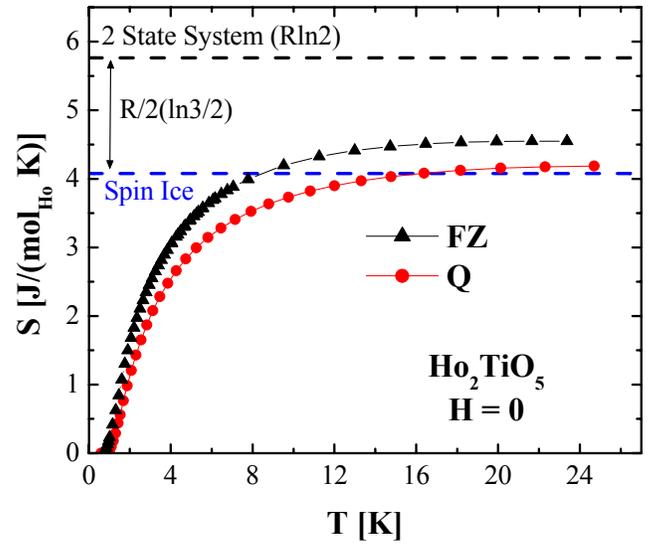